\documentclass[twocolumn,prd,amsmath,superscriptaddress]{revtex4}
\usepackage{graphicx}
\usepackage{latexsym}
\usepackage{amsmath}
\usepackage{amssymb}
\usepackage{graphics}
\usepackage{color}
\usepackage{hyperref}
\usepackage{ifthen}
\usepackage{bm}
\usepackage{color}

\definecolor{darkgreen}{RGB}{0,180,0}

\newcommand\ee{\end{equation}}
\newcommand\be{\begin{equation}}
\newcommand\eea{\end{eqnarray}}
\newcommand\bea{\begin{eqnarray}}

\def\gsim{ \lower .75ex \hbox{$\sim$} \llap{\raise .27ex \hbox{$>$}} }

\begin{document}

\title{D-BIonic Screening of Scalar Fields}
\author{Clare Burrage}
\affiliation{School of Physics and Astronomy, University of Nottingham, Nottingham, NG7 2RD, UK}%

\author{Justin Khoury}
\affiliation{Center for Particle Cosmology, University of Pennsylvania, Philadelphia, PA 19104, USA 
}

\date{\today}

\begin{abstract}
We study a new screening mechanism which is present in Dirac-Born-Infeld (DBI)-like theories. A scalar field with a DBI-like Lagrangian is minimally coupled to matter. In the vicinity of sufficiently dense sources, non-linearities in the scalar dominate and result in an approximately constant acceleration on a test particle, thereby suppressing the scalar force relative to gravity. Unlike generic $P(X)$ theories, screening happens within the regime of validity of the effective field theory, thanks to the DBI symmetry. This symmetry also allows the removal of a constant field gradient, like in galileons.  Not surprisingly, perturbations around the spherically-symmetry background propagate superluminally, but we argue for a chronology protection analogous to galileons. We derive constraints on the theory parameters from tests of gravity and discuss various extensions.
\end{abstract}

\maketitle

The discovery of cosmic acceleration has generated much activity to explore new physics associated with dark energy. 
The central motivation for ``Beyond-$\Lambda$CDM" physics is of course the smallness of the cosmological constant. But another, more
pragmatic motivation is the opportunity offered by upcoming surveys of the large scale structure, such as the Euclid mission and the Large
Synoptic Survey Telescope, which will subject the $\Lambda$CDM model to its most stringent tests to date. 

The most exciting and phenomenologically-rich possibility is if the new sector associated with dark energy (usually in the form of light scalar fields) couple to matter.
For consistency with laboratory and solar system tests of gravity, however, any such scalar field must become invisible in the local environment. This is achieved
through {\it screening mechanisms}~\cite{Jain:2010ka,Clifton:2011jh}. In regions of high density/curvature, the scalar $\phi$ develops non-linearities, which in turn decouple it from matter.

Screening mechanisms can be classified according to the nature of their non-linearities~\cite{ftnote1}:

\noindent $\bullet$~{\it $\phi$ screening.} In this case, self-interactions are governed by a potential $V(\phi)$. Whether or not the scalar develops non-linearities depends on the local value of $\phi$. This includes chameleon~\cite{Khoury:2003aq,Brax:2004qh,Gubser:2004uf}, symmetron~\cite{Hinterbichler:2010es,Olive:2007aj,Pietroni:2005pv} and varying-dilaton theories~\cite{Brax:2011ja}. Quite generally, the cosmological effects of such scalars are restricted to non-linear ($\lesssim$~Mpc) scales~\cite{Wang:2012kj}, and the lack of any symmetry makes them susceptible to radiative corrections~\cite{Upadhye:2012vh}. There are also issues with initial conditions~\cite{Erickcek:2013oma,Erickcek:2013dea}. Since $\phi$ is canonical, on the other hand, this class of theories does not suffer from superluminality issues and in principle admits a standard Wilsonian UV completion~\cite{Hinterbichler:2010wu}.

\noindent $\bullet$~{\it $\partial\phi$ screening.} These theories are of the $P(X,\phi)$ type, where $X = -(\partial\phi)^2/2$. The threshold for screening is set by the local gradient $\partial\phi$.
This includes $k$-mouflage~\cite{Babichev:2009ee} and generic $P(X)$ theories~\cite{Brax:2012jr}. 

\noindent $\bullet$~{\it $\partial^2\phi$ screening.} Non-linearities are triggered in this case by the local acceleration $\partial^2\phi$ exceeding some threshold. This mechanism originated in massive gravity~\cite{Vainshtein:1972sx} to resolve the van Dam--Veltman--Zakharov (vDVZ) discontinuity~\cite{vanDam:1970vg}. This class includes galileons~\cite{Deffayet:2001uk,Luty:2003vm,Nicolis:2008in}, which are invariant under a linear-gradient shift $\delta\phi =  c + b_\mu x^\mu$, as well as massive/resonance gravitons, {\it e.g.},~\cite{Dvali:2000hr,deRham:2010kj}.

Phenomenologically, $P(X)$ theories are similar to galileons. A key difference, however, is that the screened regime for $P(X)$ generally lies outside the regime of validity of the effective field theory (EFT). Consider $P(X) = X + \alpha X^2$, for instance. Screening requires $X^2$ to dominate over the kinetic term, at which point all sorts of higher-dimensional operators ({\it e.g.}, $X^n$) should become important.

An important exception, which is the focus of this Letter, is based on the Dirac-Born-Infeld (DBI) action
\be
{\cal L} = \Lambda^4 \sqrt{1 - \Lambda^{-4}(\partial\phi)^2}\,.
\label{DBI}
\ee
Compared to the standard DBI form, we have flipped both the overall sign of the action, together with the sign of the $(\partial\phi)^2$ coefficient. This is
necessary to achieve screening, as we will see. We will refer to the screening mechanism in this case as {\it D-BIonic screening}. This was first briefly considered in~\cite{Dvali:2010jz}. See~\cite{Mukhanov:2005bu} for cosmological applications. Expanding around $\phi = 0$, ${\cal L} = \Lambda^4 - (\partial\phi)^2/2 - (\partial\phi)^4/8\Lambda^4 + \ldots$, we see that the kinetic term has the correct sign, while the quartic term has the wrong-sign for S-matrix analyticity~\cite{Adams:2006sv}.  

The action~(\ref{DBI}) can be interpretated as the area of a (negative-tension) 3-brane moving in a 5d Minkowski space with 2 time dimensions.
The brane position is defined by $\phi(x)$, and the induced metric on the brane is
\be
\tilde{g}_{\mu\nu} = \eta_{\mu\nu} - \Lambda^{-4} \partial_\mu\phi\partial_\nu\phi \,.
\label{induced}
\ee
In terms of this metric, ${\cal L} = \Lambda^4  \sqrt{-\tilde{g}}$. The theory is therefore invariant under 5d Lorentz boosts ($\gamma \equiv 1/\sqrt{1-v^2}$):
\bea
\nonumber
\tilde{\phi}(\tilde{x}) &=& \gamma (\phi(x) + \Lambda^2 v_\mu x^\mu)\;; \\
\tilde{x}^\mu &=& x^\mu + \frac{\gamma -1}{v^2} v^\mu v_\nu x^\nu + \gamma v^\mu \frac{\phi(x)}{\Lambda^2}\,.
\label{lorentz}
\eea
This symmetry distinguishes DBI from a generic $P(X)$ theory in several important ways. Much like in DBI inflation~\cite{Silverstein:2003hf}, $\phi$
can acquire large gradients $|\vec{\nabla}\phi|\gg \Lambda^2$ while remaining within the regime of validity of the EFT, as long as higher-derivative terms remain
small. Second, a constant gradient profile can be removed by a ``boost" and is therefore unobservable, similar to galileons.
Indeed, the DBI mechanism is intermediate between $P(X)$ and the galileon.

For simplicity, we couple $\phi$ conformally to matter, ${\cal L}_{\rm coupling} = \frac{g\phi}{M_{\rm Pl}} T^\mu_{~\mu}$, where $g$ is dimensionless.
This operator breaks the symmetry~(\ref{lorentz}), but very mildly since $\Lambda\ll M_{\rm Pl}$.  In the concluding remarks, we will discuss more general matter couplings which could be relevant for cosmology. The equation of motion for $\phi$ is
\begin{equation}
\partial_{\mu}\left(\frac{\partial^{\mu}\phi}{\sqrt{1-\Lambda^{-4}(\partial\phi)^2}}\right)=-\frac{g}{M_{\rm Pl}}T^{\mu}_{\mu} \,.
\end{equation}

\noindent {\it Static profile:} Around a static point source, $T^{\mu}_{~\mu}=-\rho=-M\delta^{(3)}(\vec{x})$, the scalar field profile can be assumed static and spherically symmetric. The equation of motion 
can then be integrated once to give~\cite{Dvali:2010jz} 
\be
\frac{{\rm d}\phi}{{\rm d}r} = \frac{\Lambda^2}{\sqrt{1 + (r/r_*)^4}}\,,
\label{Esol2}
\ee
where we have introduced the $r_*$-scale
\be
r_* \equiv \frac{1}{\Lambda}\left(\frac{gM}{4\pi M_{\rm Pl}}\right)^{1/2} \,,
\label{r*}
\ee
analogously to galileon/massive gravity. Note that $r_* \sim M^{1/2}$, instead of the usual $\sim M^{1/3}$ for massive gravity.

The scalar force on a test mass is $\vec{F}_\phi=-g m\vec{\nabla}\phi/M_{\rm Pl}$. At large distances from the source ($r\gg r_*$),
$F_\phi \simeq  2g^2 F_{\rm N}$, which is just the force mediated by a massless scalar with coupling strength $g/M_{\rm Pl}$. 
This can only be made weaker than the gravitational force by tuning $g$ to be small. Close to the source, $r\ll r_*$, however, $F_\phi$ saturates to a constant, such that 
\be
\frac{F_{\phi}}{F_{\rm N}} \simeq  2g^2 \left(\frac{r}{r_*}\right)^2\,;\qquad r\ll r_*\,.
\ee
Thus, within $r_*$ the scalar force is suppressed compared to gravity, much like in the Vainshtein mechanism.

As mentioned earlier, even though $\phi'\sim \Lambda^2$ close to the source, the EFT description is still valid provided that higher-derivative terms, such
as the extrinsic curvature, remain small. The largest distance scale at which these terms become important is the Vainshtein scale~\cite{deRham:2010eu}
\be
r_{\rm UV}  \equiv \frac{1}{\Lambda}\left(\frac{gM}{4\pi M_{\rm Pl}}\right)^{1/3}  = \left(\frac{r_*^2}{\Lambda}\right)^{1/3}\,. 
\label{rUV}
\ee
The DBI term dominates on scales $r \;\gsim \; r_{\rm UV}$. There is a parametric window $r_{\rm UV} \lesssim r \lesssim r_*$ within which the EFT is valid and DBI non-linearities are important.

\vspace{0.1cm} 
\noindent {\it Blanket screening:} Despite non-linearities, the D-BIon admits an {\it exact} solution for $N$ sources. Consider two point masses $M_1$ and $M_2$, respectively located at $\vec{x} = \vec{R}$ and $\vec{x} = 0$. The coordinate system is centered around $M_2$. Defining the non-linear current $\vec{J} \equiv \frac{\vec{\nabla} \phi}{\sqrt{1 - \frac{(\partial\phi)^2}{\Lambda^4}}}$, the equation of motion can readily be integrated:
\be
\frac{\vec{J}}{\Lambda^2}  = \left(\frac{r_*^{(1)}}{|\vec{x}-\vec{R}|}\right)^2 \frac{\vec{x} - \vec{R}}{|\vec{x}-\vec{R}|} + \left(\frac{r_*^{(2)}}{r}\right)^2 \hat{x}\,,
\ee
where $r =  |\vec{x}|$ is the radial distance from $M_2$, and $\hat{x}_i \equiv \vec{x}_i/r$ is a unit vector. The current can be inverted, $\vec{\nabla} \phi =  \frac{\vec{J}}{\sqrt{1 + \frac{J^2}{\Lambda^4}}}$, to give the exact solution. We will be interested in this solution in the vicinity of $M_2$, such that $|\vec{x} - \vec{R}| \simeq R$. The result is
\be
\frac{\vec{\nabla}\phi}{\Lambda^2} \simeq \frac{ - \sqrt{1 - \frac{1}{\gamma_1^2}}  \hat{R} + \left(\frac{r_{*\,{\rm eff}}^{(2)}}{r}\right)^2 \hat{x} }{\sqrt{1 - 2 \hat{x}\cdot\hat{R} \sqrt{1 - \frac{1}{\gamma_1^2}}  \left(\frac{r_{*\,{\rm eff}}^{(2)}}{r}\right)^2 +  \left(\frac{r_{*\,{\rm eff}}^{(2)}}{r}\right)^4}}\,,
\label{exact}
\ee
where $\gamma_1 \equiv  \sqrt{ 1+  (r_*^{(1)}/R)^4}$ is the Lorentz factor due to $M_1$~\cite{ftnote2}.
We have also introduced an effective $r_*$-scale
\be
r_{*\,{\rm eff}}^{(2)} = \frac{r_{*}^{(2)}}{\sqrt{\gamma_1}}\,.
\label{r*eff}
\ee
The presence of $M_1$ shrinks the $r_*$-radius, meaning that non-linearities are pushed to shorter scales.

Far from the source, $r \gg r_{*\,{\rm eff}}^{(2)}$, we have
\be
\partial_i\phi  \simeq - \Lambda^2\sqrt{1 - \gamma_1^{-2}} \hat{R}_i + \frac{gM_2}{4\pi \gamma_1 M_{\rm Pl}} P_{ij} \hat{x}^j \,,
\label{linearfield}
\ee
where $P_{ij} = \delta_{ij}  + \left(\gamma_1^{-2} - 1 \right) \hat{R}_i \hat{R}_j $.
The first term is just the profile due to $M_1$. The second term represents the $M_2$
contribution, distorted by the presence of $M_1$. For $\gamma_1 > 1$, the presence of $M_1$
suppresses the field gradient due to $M_2$. We can read off that the effective coupling is least suppressed
in the plane orthogonal to the $M_1$-$M_2$ axis, $g_{\rm eff}^\perp = g/\gamma_1$. Along the $M_1$-$M_2$ axis, 
$g_{\rm eff}^{||} = g/\gamma_1^3$.

Close to the source, $r \ll r_{*\,{\rm eff}}^{(2)}$,~\eqref{exact} reduces to
\be
\partial_i\phi  \simeq \Lambda^2\hat{x}_i\,.
\ee
This is just the standard non-linear profile for $M_2$. The presence of $M_1$ is irrelevant.

\vspace{0.1cm}
\noindent {\it Solar System Constraints:} We now turn to constraints on the theory from solar system tests of gravity. As with galileons~\cite{Dvali:2002vf,Dvali:2007kt,Afshordi:2008rd}, 
the most stringent constraint comes from the Earth-Moon system. As we will see, blanket screening from the Sun is important, with $\gamma_\odot \simeq (r_*^\odot/{\rm AU})^2 \gg 1$.
From~\eqref{r*eff}, the renormalized $r_*$ for the Earth becomes independent of $\Lambda$ in this limit:
\be
r_{*\,{\rm eff}}^\oplus \simeq \sqrt{\frac{M_\oplus}{M_\odot}} R \simeq 2\times 10^5~{\rm km}\,,
\ee
where we have substituted $M_\oplus \simeq 6\times 10^{24}~{\rm kg}$, $M_\odot \simeq 2\times 10^{30}~{\rm kg}$, and $R  = 1~{\rm AU} \simeq 10^8~{\rm km}$.
Although $r_{*\,{\rm eff}}^\oplus$ is only half the Earth-Moon distance, $d_{\rm Earth-Moon} \simeq 4\times 10^5~{\rm km}$, let us for simplicity approximate the
Earth's profile by the linear expression~\eqref{linearfield}. The maximum force is orthogonal to the Earth-Sun axis, and is given by
\be
F_\varphi^{\rm Earth-Moon}  \simeq F_{\rm N}\frac{g^2}{\gamma_\odot}\,.
\ee
Lunar Laser Ranging (LLR) observations require~\cite{Dvali:2002vf,Dvali:2007kt,Afshordi:2008rd}
\be
\frac{g^2}{\gamma_\odot} \simeq \left(\frac{gR}{r_*^\odot}\right)^2 \lesssim 2.4\times 10^{-11}\,,
\label{LLR}
\ee
which translates to
\be
\sqrt{g}\Lambda \lesssim 4\times 10^{-5}~{\rm eV}\,.
\ee
This is the main constraint on D-BIonic parameters. (Incidentally, for $g\sim {\cal O}(1)$,~\eqref{LLR} implies $\gamma_\odot \;\gsim\; 4\times 10^{10}$,
thus confirming the importance of blanket screening from the Sun. One can also easily check that blanket screening from the Milky Way galaxy, ignored in this calculation, is indeed a negligible effect.)

It is useful to express various quantities in terms of this bound. The $r_*$ scale for the Sun can be written as
\be
r_*^\odot = 3\times 10^{13} g \left[\frac{4\times 10^{-5}~{\rm eV}}{\sqrt{g}\Lambda}\right]~{\rm km}\,.
\ee
For $g\sim {\cal O}(1)$, this is much larger than the solar system.

As mentioned earlier, the D-BIon solution breaks down at the scale $r_{\rm UV}$ where other DBI Galileon terms would kick in. For the Sun,~\eqref{rUV} gives
\be
r_{\rm UV}^\odot = 2 \times 10^{7} g^{5/6} \left[\frac{4\times 10^{-5}~{\rm eV}}{\sqrt{g}\Lambda}\right]~{\rm km}\,.
\ee
Since $r_{\rm UV}^\odot$ is larger than the Sun's radius, we cannot describe light-bending constraints within the D-BIon effective theory. Similarly for the Earth,
\be
r_{\rm UV}^\oplus = 2 \times 10^{5} g^{5/6} \left[\frac{4\times 10^{-5}~{\rm eV}}{\sqrt{g}\Lambda}\right]~{\rm km}\,.
\ee
This is comparable to the Earth-Moon distance, which means that the LLR analysis above is barely within the effective description.
It also implies that gravity constraints from laboratory, geophysical and satellite tests are not describable within the D-BIon effective theory.
This can be viewed as strong motivation for including the higher-order DBI Galileon terms, which would kick in at $r_{\rm UV}$. We are currently investigating
this~\cite{future}.

\vspace{0.1cm}
\noindent {\it Superluminality and Chronology Protection:} Analogously to galileons, perturbations on certain backgrounds can propagate superluminally. This is usually interpreted as a sign that the UV completion is non-standard~\cite{Adams:2006sv}. However, recent developments in galileon dualities~\cite{Bellucci:2002ji,Creminelli:2013fxa,deRham:2013hsa,Creminelli:2014zxa,deRham:2014lqa} shows that the issue of superluminality is at the very least more subtle than anticipated. We will take a conservative point of view and, for consistency, argue for chronology protection: closed time-like curves (CTCs) cannot form within the regime of validity of the EFT. 

Consider the quadratic action for small perturbations around some background,
\be
\mathcal{L}_\varphi = -\frac{1}{2}G^{\mu\nu}\partial_{\mu}\varphi\partial_{\nu}\varphi  \,,
\label{pertaction}
\ee
where the effective metric is $G^{\mu\nu}= \gamma g^{\mu\nu}+\frac{\gamma^3}{\Lambda^4}\partial^{\mu}\phi\partial^{\nu}\phi$. 
For perturbations around a spherically-symmetric profile sourced by a mass $M$, the effective metric (with lowered indices) in spherical coordinates is
\be
G_{\mu\nu}= \gamma^{-1} {\rm diag}\left(-1, \gamma^{-2},r^2,r^2\sin^2\theta\right)\,.
\ee
Suppressing the angular directions, the null cones for radial propagation are defined by
\be
v_\pm \sim \left(\begin{array}{c}
1\\
\pm \gamma
\end{array}\right)\,.
\end{equation}
The null cone gets wider as $\gamma$ increases, allowing for superluminal propagation radially. Note, however, that the null-cone never tips over to allow propagation to negative coordinate times.  There always exists a set of surfaces which are space-like with respect to both $G_{\mu\nu}$ and $\eta_{\mu\nu}$, and these can be chosen as Cauchy surfaces on which to specify initial conditions.  Therefore causal propagation is always possible, as in k-essence theories~\cite{Babichev:2007dw}.

The reader may be concerned that  Lorentz boosting the system would allow travel to negative times, and this is indeed the case. Under the Lorentz transformation 
$t= \Gamma (t^{\prime}+vr^{\prime})$, $r= \Gamma (r^{\prime}+vt^{\prime})$, with $\Gamma \equiv 1/\sqrt{1-v^2}$ (to avoid confusion with $\gamma$), 
the $v_\pm $ null vectors become 
\be
v_\pm \sim \left(\begin{array}{c}
 \Gamma^{-2} \mp  v \gamma \left(1-\gamma^{-2}\right)  \\
 \pm\gamma \left(1 -\frac{v^2}{\gamma^2}\right)
\end{array}\right) \,,
\ee
where  $\gamma$ is now a function of both the Lorentz transformed coordinates $\gamma\equiv \gamma(r^{\prime}+vt^{\prime})$.
Deep inside the screening radius where $\gamma\gg 1$, the null cones tip over to include negative coordinate times when
\be
\gamma > \frac{1}{v\Gamma^2} \,.
\ee
It seems naively possible for a scalar fluctuation to propagate causally to some point $t^{\prime}<0$, interact with a Standard Model particle moving on geodesics of $\eta_{\mu\nu}$, and then for this particle to propagate forward in time, thereby creating a CTC.  However it is straightforward to check that the Hamiltonian for scalar fluctuations diverges as $\gamma \rightarrow 1/v\Gamma^2$, hence it is not possible for the system to evolve from a situation in which no CTCs exist to one in which such a curves are formed, while remaining within the regime of validity of the EFT. The divergence of the scalar Hamiltonian at the point of formation of closed time-like curves was first demonstrated for Galileons in~\cite{Burrage:2011cr}, and is an extension of HawkingÕs Chronology Protection Conjecture~\cite{Hawking:1991nk}.

\vspace{0.1cm}
\noindent {\it Conclusions:} In this paper we have explored a novel screening mechanism based on the DBI action, focusing primarily on constraints from local tests of gravity. Interestingly, the scale $\Lambda$ is constrained to be of order the dark energy scale, which is promising for cosmology. A detailed analysis of the cosmological evolution is currently underway and will appear elsewhere~\cite{future}. In particular, it would be interesting to study the impact of D-BIons on halo profiles and cluster/galactic dynamics. Various extensions of the theory suggest themselves. The DBI Lagrangian~(\ref{DBI}) is only one of five DBI galileon terms, and it will be interesting to generalize the mechanism to include such terms as well~\cite{future}. The matter coupling can also be generalized. A more natural choice is to couple matter fields to a generalization of the brane metric~(\ref{induced}), $\tilde{g}_{\mu\nu} = e^{2g\phi/M_{\rm Pl}}\left(\eta_{\mu\nu} - \Lambda^{-4} \partial_\mu\phi\partial_\nu\phi\right)$. This will have no impact on the static profiles studied here, but can play an important role for the cosmological evolution~\cite{Koivisto:2013fta}.

\smallskip{\em Acknowledgments.}   We thank L.~Berezhiani, P.~Creminelli and D.~Stefanyszyn for helpful discussions. We are especially grateful to C.~de Rham for correcting an error in an earlier version and for enlightening discussions. J.K. is supported by NSF CAREER Award PHY-1145525 and NASA ATP grant NNX11AI95G. C.B. is supported by a Royal Society University Research Fellowship.


\end{document}